\documentclass[12pt]{article}
\usepackage{amsmath}
\usepackage{graphicx}
\usepackage{color}
\begin{document}
\baselineskip=18 pt 
\begin{center}
{\large{\bf  Cylindrical symmetric, non-rotating and non-static or static black hole solutions and the naked singularities}}
\end{center} 

\vspace{.5cm}

\begin{center}
{\bf Faizuddin Ahmed}\footnote{faizuddinahmed15@gmail.com ; faiz4U.enter@rediffmail.com}\\
{\it Ajmal College of Arts and Science, Dhubri-783324, Assam, India}
\end{center}

\vspace{.5cm}

\begin{abstract}

In this work, a four-dimensional cylindrical symmetric and non-static or static space-times in the backgrounds of anti-de Sitter (AdS) spaces with perfect stiff fluid, anisotropic fluid and electromagnetic field as the stress-energy tensor, is presented. For suitable parameter conditions in the metric function, the solution represents non-static or static non-rotating black hole solution. In addition, we show for various parameter conditions, the solution represents static and/or non-static models with a naked singularity without an event horizon.

\end{abstract}

\vspace{0.2cm}


\vspace{.5cm}

\section{Introduction}

In General Relativity, it is a subject of long-standing interest to look for the exact solutions of Einstein's field equations. Among these exact solutions, black hole solutions take an important position because thermodynamics, gravitational theory, and quantum theory are connected in quantum black hole physics. In addition, black holes might play an important role in developing a satisfactory full quantum theory of gravity which does not exist till today. Black holes are regions of space-time from which nothing, not even light, can escape. A typical black hole is the result of the gravitational force becoming so strong that one would have to travel faster than light to escape its pull. Such black holes generically contain a space-time singularity at their center ; thus we cannot fully understand a black hole without also understanding the nature of singularities. Black holes, however, raise several additional conceptual problems and questions on their own. When quantum effects are taken into account, black holes, although they are nothing more than regions of space-time, appear to become thermodynamic entities, with a temperature and an entropy. The fundamental features of a black hole is the topology. In four-dimensional asymptotically flat stationary space-time, Hawking showed that a black hole has necessarily a $S^{2}$ topology, provided that the dominant energy condition holds \cite{H}. This result extends to outer apparent horizons in black hole space-times that are not necessarily stationary \cite{H2}. Such restrictive uniqueness theorems do not hold in higher dimensions, the most famous counter-example being the black ring of Emparan {\it et al} \cite{Emp}, with horizon topology $S^2\times S^1$. Nevertheless, Galloway {\it et al} \cite{Gall} have shown that in arbitrary dimension, cross sections of the event horizon (in the stationary case) and outer apparent horizons (in the general case) admit metrics of positive scalar curvature. Hawking's theorem was later generalized by Gannon \cite{Gannon} who had proved, replacing stationary by some weaker assumptions, that the horizon must be either spherical or toroidal. The topological censorship theorem \cite{Friedmann} also plays an important role in black hole physics which states that in asymptotically flat space-times, only spherical horizons can give rise to well defined causal structure for a black hole. This is circumvented by the presence of a negative cosmological constant, in which case well defined black holes with locally flat or hyperbolic horizons have been shown to exist \cite{JPS,L,DR}. This kind of black holes with topologically non-trivial anti-de Sitter (AdS) asymptotics are relevant in testing the AdS/CFT correspondence.

In the framework of four-dimensions, it is well known that generic black hole solutions to Einstein-Maxwell equations are Kerr-Newman solutions \cite{Kerr,Kerr2}, which are characterized by only three parameters : the mass (M), charge (q), and angular momentum (J). It is often referred to as the non-hair conjecture of black holes and the space-time is asymptotically flat. When a non-zero cosmological constant is introduced, the space-time will become asymptotically de Sitter (dS) or anti-de Sitter (AdS) spaces depending on the sign of the cosmological constant. In a recent paper \cite{Lemos}, Lemos constructed a cylindrical symmetric rotating and/or non-rotating and static black hole solutions (black strings) in four-dimensional Einstein gravity with a negative cosmological constant. The black string solutions of Lemos is asymptotically AdS not only in the transverse directions, but also in the string direction. Huang {\it et al} \cite{Huang} further constructed the so-called torus-like black holes with the topology $R^2\times S^1\times S^1$. The topological black holes in four-dimensional non-static space-time with non-zero cosmological constant, were investigated in \cite{EPJA}. Y. Wu {\it et al} \cite{Wu} constructed a topologically charged static black hole solution in four-dimensions. Once the energy condition is relaxed, a black hole can have quite different topology. Such examples can occur even in four-dimensional space-times where the cosmological constant is negative \cite{Cai2,Mann,Brill,Cai3,Klemm,Klemm2}.

The curvature singularities of matter-filled as well as vacuum space-times are recognized from the divergence of the energy-density and/or the scalar curvature invariant, such as the zeroth order scalar curvature $R_{\mu\nu\rho\sigma}\,R^{\mu\nu\rho\sigma}$ (so called the Kretschmann scalar), and $R_{\mu\nu\rho\sigma}\,R^{\rho\sigma\lambda\tau}\,R^{\mu\nu}_{\lambda\tau}$. In addition, by analysing the outgoing radial null geodesics of a space-time containing space-time singularity, one can determine whether the curvature singularity is naked or covered by an event horizon (see Refs.\cite{Jos,Jos1} for detail discussion). For the strength of curvature singularities, two conditions are generally used : first one being the {\it strong curvature condition} (SCC) given by Tipler \cite{Tipler}, and second one is the {\it limiting focusing condition} (LFC) given by Krolak \cite{Kro}. Meanwhile, the curvature singularities are basically of three types : first one being a space-like singularity ({\it e. g.} Schwarzschild singularity), second one is the time-like singularity, and third one is the null singularity. In time-like singularity, two possibilities arise : $(i)$ there is an event horizon around a time-like singularity ({\it e. g.} RN black holes). Here an observer cannot see a time-like singularity from an outside region of the space-time, that is, the singularity is covered by an event horizon ; $(ii)$ there is no event horizon around a time-like singularity which we called naked singularities (NS), and it would observable for far away observers. Therefore existence of a naked singularity would present opportunities to observe the physical effects near the very dense regions that formed in the very final stages of a gravitational collapse. But in a black hole scenario, such regions are necessarily hidden within the event horizon. Attempts have been made to provide the theoretical framework to devise a technique to distinguish between black holes and naked singularities from astrophysical data mainly through gravitational lensing (GL) method. Some significant works in this direction are the study of strong gravitational lensing in the Janis-Newman-Winicour space-time \cite{KS,KS2}, and its rotating generalization \cite{Yaz}, and notable work in \cite{Wer,Bambi,Bambi2,Hioki}. Other workers have shown that naked singularities and black holes can be differentiated by the properties of the accretion disks that accumulate around them. Consequently, the study of naked singularities and space-time with such objects are of considerable current interest. In \cite{Chow}, the authors have enumerated three possible end states of gravitational collapse. There are examples of gravitational collapse model which formed a naked curvature singularity known. The earliest model is the Lemaitre-Tolman-Bondi (LTB) \cite{Lei,Tol, Bondi} solutions, a spherically symmetric in-homogeneous collapse of dust fluid that admits both naked and covered singularity. Papapetrou \cite{pap} pointed out the formation of naked singularities in Vaidya \cite{vai} radiating solution, a null dust fluid space-time generated from Schwarzschild vacuum solution. The examples of spherically symmetric gravitational collapse space-times with naked singularities would be \cite{Christ,Des,Gov,Brave,Clarke,Jos3,Jos4,Glass,Rocha,Herra,Kras} and many more. To counter the occurrence of naked singularity in a solution of the Einstein's field equations, R. Penrose proposed a Cosmic Censorship Conjecture (CCC) \cite{Pen1,Pen2,Pen3}. However, the general and/or detail proofs of this Conjecture has not yet been given. On the contrary, there is no mathematical details yet known which forbid the appearance of naked singularity in a solution of the field equations.

The Gravitational collapsing of cylindrically symmetric models that is formed a naked singularity has been discussed in \cite{Thor,Hay}. Apostolatos {\it et al} \cite{Apo} investigated the collapse of counter-rotating dust shell cylinder. Echeverria \cite{Eche} has studied the evolution of cylindrical dust shell analytically at late times and numerically for all times. Guttia {\it et al} \cite{Gutt} have studied the collapse of non-rotating, infinite dust cylinders. Nakao {\it et al} \cite{Nakao} have studied the high-speed collapse of cylindrically symmetric thick shell composed of dust, and perfect fluid with non-vanishing pressure \cite{Nakao1}. Recent work describes the cylindrically symmetric collapse of counter-rotating dust shell \cite{Gonc,Nolan,Peri}, self-similar scalar field \cite{Wang}, axially symmetric vacuum space-time \cite{Faiz}, cylindrically symmetric anisotropic fluid space-time \cite{Faiz2}, axially symmetric null dust space-time \cite{Faiz3}, cylindrically symmetric type N anisotropic and null fluid space-time \cite{Faiz4}, cylindrically symmetric vacuum space-time \cite{Faiz5}, and cylindrically symmetric and static anisotropic fluid space-time \cite{Faiz6}. Some other examples of non-spherical gravitational collapse with naked singularity would be discussed \cite{Chi,Piran,Th,Morgan,Lete,JMM,Bondi2,M,M2,Wang2}.

The Einstein field equations with cosmological constant $\Lambda$ are given by
\begin{equation}
G_{\mu\nu}+\Lambda\,g_{\mu\nu}=\kappa\,T_{\mu\nu},
\label{field}
\end{equation}
where $T_{\mu\nu}$ is the stress-energy tensor. For our present work, we have chosen the following stress-energy tensors:

(i) Perfect fluid :
\begin{equation}
T^{0}_{0}=-\rho,\quad T^{1}_{1}=T^{2}_{2}=T^{3}_{3}=p,
\label{perfect}
\end{equation}
where $\rho$ is the energy-density and $p$ is the isotropic pressure.

(ii) Electromagentic field :
\begin{equation}
T_{\mu\nu}=\frac{2}{\kappa}\,(g^{\alpha\beta}\,F_{\alpha\mu}\,F_{\beta\nu}-\frac{1}{4}\,g_{\mu\nu}\,F_{\alpha\beta}\,F^{\alpha\beta}).
\label{em}
\end{equation}
where $F_{\mu\nu}=\partial_{\mu} A_{\nu}-\partial_{\nu} A_{\mu}$ is the electromagentic field tensor satisfying the following conditions
\begin{equation}
F^{\mu\nu}_{\,\,\,;\nu}=0,\quad F_{[\mu\nu;\alpha]}=0,
\label{}
\end{equation}
for the source-free regions, where $A_{\mu}$ is the four-vector potential.

(ii) Anisotropic fluid :
\begin{equation}
T^{0}_{0}=-\rho,\quad T^{1}_{1}=p_{r},\quad T^{2}_{2}=p_{t}=T^{3}_{3},
\label{anisotropic}
\end{equation}
where $\rho$ is the energy-density, and $p_{r},p_{t}$ are the radial and tangential pressures, respectively.

In the present work, taking into account the energy conditions we attempt to generalize a four-dimensional cylindrically symmetric black hole into non-static case which may represent model of black holes and/or naked singularities under various parameters condition.

\section{A four-dimensional non-static space-times}

The four dimensional time-dependent cylindrical symmetry space-time is given by
\begin{eqnarray}
ds^2&=&g_{\mu\nu}\,dx^{\mu}\,dx^{\nu}\nonumber\\
&=&-f(r)\,dt^2+\frac{1}{f(r)}\,dr^2+r^2\,(A\,d\phi^2+\frac{1}{A}\,dz^2),
\label{1}
\end{eqnarray}
where  the ranges of the coordinates are $-\infty < t < \infty$, $r \in [0, +\infty[$, $-\infty < z < \infty$, $\phi \in [0, 2\,\pi)$ and
\begin{equation}
f(r)=(\frac{2\,\beta}{r}+\frac{\alpha^2}{r^2}+b^2\,r^2+\delta\,r),\quad A=A(t)=e^{-2\,\gamma\,t}.
\label{2}
\end{equation}
Here $\beta$ represents mass parameter, $\alpha$ represents charge, $b^2$ is related to cosmological constant ($\Lambda$), $\delta$ and $\gamma$ are integer. The metric functions with its inverse are
\begin{eqnarray}
g_{00}&=&-f(r)=\frac{1}{g^{00}},\quad g_{11}=\frac{1}{f(r)}=\frac{1}{g^{11}},\nonumber\\
g_{22}&=&r^2\,e^{-2\,\gamma\,t}=\frac{1}{g^{22}},\quad g_{33}=r^2\,e^{2\,\gamma\,t}=\frac{1}{g^{33}}.
\label{3}
\end{eqnarray}
The metric is Lorentzian with signature $(-,+,+,+)$ and the determinant of the corresponding metric tensor $g_{\mu\nu}$ is
\begin{equation}
det\;g=-r^4.
\label{4}
\end{equation}
The non-zero components of the Einstein tensor $G^{\mu\nu}$ are
\begin{eqnarray}
G^{0}_{0}&=&3\,b^2-\frac{\alpha^2}{r^4}+\frac{2\,\delta}{r}+\frac{\gamma^2}{f(r)},\nonumber\\
G^{1}_{1}&=&3\,b^2-\frac{\alpha^2}{r^4}+\frac{2\,\delta}{r}-\frac{\gamma^2}{f(r)},\nonumber\\
G^{2}_{2}=G^{3}_{3}&=&3\,b^2+\frac{\alpha^2}{r^4}+\frac{\delta}{r}-\frac{\gamma^2}{f(r)},
\label{5}
\end{eqnarray}

The time-like four-velocity vector when $t$ is time-like and $r$ is space-like defined by
\begin{equation}
u^{\mu}={f(r)}^{-1/2}\,\delta^{\mu}_{t},\quad u^{\mu}\,u_{\mu}=-1,
\label{6}
\end{equation}
such that the four-acceleration vector
\begin{equation}
a^{\mu}=u^{\mu\,;\,\nu}\,u_{\nu}=k(r)\,\delta^{\mu}_{r},\quad k(r)=\frac{1}{2}\,f'(r).
\label{7}
\end{equation}

Before proceeding, we do the following transformation 
\begin{equation}
t\rightarrow v-\int \frac{dr}{f(r)}
\label{8}
\end{equation}
into the metric (\ref{1}), one will get
\begin{equation}
ds^2=-f(r)\,dv^2+2\,dv\,dr+r^2\,(H(v,r)\,d\phi^2+\frac{1}{H(v,r)}\,dz^2),
\label{9}
\end{equation}
where
\begin{equation}
H(v,r)=e^{-2\,\gamma\,v}\,e^{2\,\gamma\,\int \frac{dr}{f(r)}}.
\label{10}
\end{equation}

Now, we study the following cases of the above space-time in details.

\vspace{0.1cm}
\begin{center}
{\bf Case 1} : $\beta\rightarrow -\beta$,\quad $b^2>0$,\quad $\alpha=0$,\quad $\delta=0$,\quad $\gamma\neq 0$.
\end{center}
\vspace{0.1cm}

The function $f(r)$ and the quantity $k(r)$ under this case are
\begin{equation}
f(r)=(b^2\,r^2-\frac{2\,\beta}{r})=\frac{2\,\beta}{r}\,(\frac{r^3}{r^{3}_{0}}-1),\quad k(r)=(b^2\,r+\frac{\beta}{r^2}).
\label{11}
\end{equation}
The quantity $k(r)$ for $f(r_0)=0$ is $k(r=r_0)=\frac{3}{2}\,b^2\,r_0>0$ which is space-like where, $r=r_0=(\frac{2\,\beta}{b^2})^{\frac{1}{3}}$. Here $v$ is time-like as long as $r > r_0$ and $k(r)$ is space-like for all $r>0$. One can easily check that the scalar curvature invariant constructed from the Riemann tensor is singular at $r=0$ which is covered by an event horizon $r=r_0$ and vanish rapidly at $r\rightarrow \infty$.

Consider the stress-energy tensor the perfect fluid (\ref{perfect}), from the field equations (\ref{field}) using (\ref{5}), we get
\begin{equation}
\Lambda=-3\,b^2<0,\quad \kappa\,\rho=\kappa\,p=-\frac{\gamma^2}{\frac{2\,\beta}{r}\,(\frac{r^3}{r^{3}_{0}}-1)},
\label{13}
\end{equation}
The energy-density violate the weak energy conditions in the exterior region $ r > r_0$. Inside the black hole region $0 < r < r_0$, the energy-density satisfy the energy conditions. If one takes $\gamma=0$, this corresponds to the well-known Lemos cylindrical symmetric and static black hole solution \cite{Lemos}.

Thus from the above analysis we have seen that for $\gamma \neq 0$, the solution represent a non-static stiff fluid black holes model in the background of anti-de Sitter (AdS) spaces with black holes region $r \leq r_0$, a generalization of Lemos static black hole solution into non-static one.

\vspace{0.1cm}
\begin{center}
{\bf Case 2} : $\beta=0$,\quad $\alpha^2\rightarrow -\alpha^2$, \quad $b^2>0$,\quad $\gamma=0$,\quad $\delta=0$.
\end{center}
\vspace{0.1cm}

The function $f(r)$ and the surface gravity $k (r)$ are given by
\begin{eqnarray}
&&f(r)=(b^2\,r^2-\frac{\alpha^2}{r^2})=\frac{\alpha^2}{r^2}\,(\frac{r^4}{r^{4}_{0}}-1),\nonumber\\
&&k (r)=(b^2\,r+\frac{\alpha^2}{r^3}),\quad r_0=(\frac{\alpha^2}{b^2})^{\frac{1}{4}}.
\label{nnn1}
\end{eqnarray}
The Krestchmann scalar given by
\begin{equation}
R_{\mu\nu\rho\sigma}\,R^{\mu\nu\rho\sigma}=\frac{56\,\alpha^4}{r^8}+24\,b^4
\label{nnn2}
\end{equation}
\textcolor{blue}{is singular at $r=0$} which is covered by an event horizon $r=r_0$ and becomes finite at $r\rightarrow \infty$. The surface gravity for $f(r_0)=0$ is $k (r_0)=2\,b^2\,r_0>0$ which is space-like. Here $t$ is time-like and $r$ is space-like for $r>r_0$ and vice-versa in the region $r <r_0$.

Considering anisotropic fluid (\ref{anisotropic}) as the stress-energy tensor, from the field equations (\ref{field}) using (\ref{5}) we have
\begin{equation}
\Lambda=-3\,b^2,\quad \kappa\,\rho=\kappa\,p_{t}=-\frac{\alpha^2}{r^4},\quad \kappa\,p_{r}=\frac{\alpha^2}{r^4}.
\label{nnn3}
\end{equation}
We have seen from above that at the point of curvature singularity $r=0$, the stress-energy tensor diverge and vanish rapidly at $r\rightarrow \infty$. Thus the solution is radially asymptotically AdS. For stationary black hole solution case, we have from (\ref{9})
\begin{equation}
ds^2=-f(r)\,dv^2+2\,dv\,dr+r^2\,(d\phi^2+dz^2).
\label{nnn4}
\end{equation}

For light-like curves, from the metric (\ref{nnn4}) the null condition implies (radial condition $\phi=const$, $z=const$)
\begin{equation}
-f(r)\,dv+2\,dr=0\Rightarrow 2\,\frac{dr}{dv}=f(r).
\label{n1}
\end{equation}
\quad Thus we see that
\begin{eqnarray}
\frac{dr}{dv}=\frac{1}{2}\,f(r)=\frac{\alpha^2}{r^3}\,(\frac{r^4}{r^{4}_{0}}-1) &>& 0\quad \mbox{for}\quad r >r_0\nonumber\\
&<& 0\quad \mbox{for}\quad r < r_0.
\label{n2}
\end{eqnarray}

Let us consider a closed two-surface, $S$, of constant $v$ and $r$, from the metric (\ref{nnn4}) we have two null vector fields
\begin{eqnarray}
k^{\mu}&=&-\partial_{r}\quad \mbox{(the inner null normal)}\nonumber\\
l^{\mu}&=&\partial_{v}+\frac{1}{2}\,f(r)\,\partial_{r}\,\mbox{(the outer null normal)}.
\label{14}
\end{eqnarray}
We compute the expansion scalar associated with ${\bf k, l}$ for two-surface \cite{Nielsen,Valerio}
\begin{eqnarray}
\Theta_{k}&=&h^{\mu\nu}\,\nabla_{\nu}\,k_{\mu}=-\frac{2}{r}\nonumber\\
\Theta_{l}&=&h^{\mu\nu}\,\nabla_{\nu}\,l_{\mu}=\frac{f(r)}{r}=\frac{\alpha^2}{r^3}\,(\frac{r^4}{r^{4}_{0}}-1),
\label{15}
\end{eqnarray}
where $h_{\mu\nu}$ is the spatial metric in the two-space defined by
\begin{equation}
h_{\mu\nu}=g_{\mu\nu}+k_{\mu}\,l_{\nu}+l_{\mu}\,k_{\nu},
\label{16}
\end{equation}
satisfy the conditions $h_{\mu\nu}\,k^{\nu}=h_{\mu\nu}\,l^{\nu}=0$, $h^{\mu}_{\mu}=2$ and $h^{\mu}_{\nu}$ is the projection operator.

Once we have the expansions, following Penrose definition \cite{Roger-Penrose} one can define that the two-surface, $S$, of constant $v$ and $r$ is {\it trapped}, {\it marginally trapped}, or {\it un-trapped}, according to whether $\Theta_{l}\,\Theta_{k}>0$, $\Theta_{l}\,\Theta_{k}=0$, or $\Theta_{l}\,\Theta_{k}<0$. An {\it apparent horizon}, or {\it trapping horizon} in Hayward's terminology \cite{Hay,Hay2} is defined as a hypersurface foliated by {\it marginally trapped} surfaces. However, this need not be the case in regions of strong curvature.

In our case, we have
\begin{eqnarray}
\Theta_{l} &>&0\quad \mbox{for} \quad r > r_0\nonumber\\
&=&0\quad \mbox{for} \quad r = r_0\nonumber\\
&<&0\quad \mbox{for} \quad r < r_0
\label{17}
\end{eqnarray}
\quad Thus, in the quotient manifold with metric $g=-f(r)\,dv^2+2\,dv\,dr$, we have 
\begin{eqnarray}
\mbox{Regular region}, {\it R}&=&\{(v,r): r >r_0\},\nonumber\\
\mbox{Trapped region}, {\it T}&=&\{(v,r): r <r_0\},\nonumber\\
\mbox{Apparent region}, {\it A}&=&\{(v,r): r =r_0\}.
\label{18}
\end{eqnarray}
The induced metric on {\it A} is
\begin{equation}
h=2\,dv\,dr=2\,(\frac{1}{2}\,f(r)\,dv)\,dv=f(r)\,dv^2,
\label{19}
\end{equation}
vanish at $r=r_0$, a null hypersurface condition, where we have used Eq. (\ref{n1}). So {\it A} is a null hypersurface.

The 2-surface $S$ with $\Theta_{k}<0$ \cite{Nielsen,Valerio,Roger-Penrose,Hay2,Blau}
\begin{eqnarray}
S\quad \mbox{is called untrapped} \quad \mbox{if} \quad \Theta_{l}&>&0\nonumber\\
\quad\mbox{marginally trapped} \quad \mbox{if} \quad \Theta_{l}&=&0\nonumber\\
\quad\mbox{trapped} \quad \mbox{if} \quad \Theta_{l}&<&0
\label{20}
\end{eqnarray}
and we can rephrase as
\begin{eqnarray}
S_{v,r}\quad \mbox{is untrapped} \quad \mbox{for} \quad r &>& r_0\nonumber\\
\quad\mbox{marginally trapped} \quad \mbox{for} \quad r &=& r_0\nonumber\\
\quad\mbox{trapped} \quad \mbox{for} \quad r &<& r_0
\label{21}
\end{eqnarray}
A {\it future outer trapping horizon} (FOTH) is the closure of a surface foliated by {\it marginal surfaces}, such that $k^{\mu}\,\nabla_{\mu}\,\Theta_{l}|_{r=r_0} < 0$, $\Theta_{l}=0$ and $\Theta_{k}<0$. In our case, we have
\begin{eqnarray}
k^{\mu}\,\nabla_{\mu}\,\Theta_{l}&=&-\alpha^2\,(\frac{3}{r^4}+\frac{1}{r^{4}_{0}})<0,\nonumber\\
k^{\mu}\,\nabla_{\mu}\,\Theta_{l}|_{r=r_0}&=&-\frac{4\,\alpha^2}{r^{4}_{0}}=-4\,b^2=\frac{4}{3}\,\Lambda<0.
\label{22}
\end{eqnarray}

To show that the region $ r \leq r_0$ represent a black hole region for the metric (\ref{1}) under this case with the event horizon exist at $r=r_0$, we apply the theorem (theorem 1.2.5) \cite{PT}. From (\ref{nnn4}), we get
\begin{equation}
g_{vv}=-f(r),\quad g_{\phi\phi}=r^2,\quad g_{vr}=1=g_{rv},\quad g_{zz}=r^2.
\label{23}
\end{equation}
The inverse metric tensor for the metric (\ref{nnn4}) in this case can be express as
\begin{equation}
g^{\mu\nu}\,\partial_{\mu}\,\partial_{\nu}=2\,\partial_{v}\,\partial_{r}+f(r)\,\partial_{r}^2+r^{-2}\,\partial_{\phi}^2+r^{-2}\,\partial_{z}^2.
\label{24}
\end{equation}
Since the metric component $g^{vv}$ from (\ref{nnn4}) is
\begin{equation}
g^{vv}=0\Rightarrow g(\nabla v,\nabla v)=0,
\label{25}
\end{equation}
which implies that the integral curves of 
\begin{equation}
\nabla v=g^{\mu\nu}\,\partial_{\mu}\,v\,\partial_{\nu}=\partial_{r},
\label{26}
\end{equation}
are null, affinely parameterized geodesic, that means, the curves are the null geodesic (since the conditions $X^{\nu}\,\nabla_{\nu}\,X^{\mu}=0$ hold where, we have defined $X=\nabla\,v$).

Again from metric (\ref{nnn4}), we have
\begin{equation}
\nabla r=g^{\mu\nu}\,\partial_{\mu}\,r\,\partial_{\nu}=\partial_{v}+f(r)\,\partial_{r}=\partial_{v}
\label{27}
\end{equation}
provided $f(r)=0\Rightarrow r=r_0=(\frac{\alpha^2}{b^2})^{\frac{1}{4}}$. Since at $r=r_0$, the metric function $f(r)=0$ which implies $g^{rr}=f(r=r_0)=0$, the null hypersurface condition. Therefore the curve $\gamma(\lambda)$ defined by $\{v=v(\lambda), r=r_0, \phi=const, z=const\}$ are the null geodesics where, $\lambda$ is an affine parameter. The null geodesics are the generators of the event horizon.

Now we study the geodesic motion of test particles. The Lagrangian of the metric (\ref{1}) in the constant $z-planes$ in this case is given by
\begin{equation}
    -f (r)\,\dot{t}^2+\frac{\dot{r}^2}{f (r)}+r^2\,\dot{\phi}^2=-\epsilon,
    \label{geodesics}
\end{equation}
where $\epsilon=0,1$ for photon and massive particles, respectively. There are two constants of motion which are given by
\begin{equation}
    \dot{t}=\frac{E}{f (r)},\quad \dot{\phi}=\frac{L}{r^2}.
    \label{constants}
\end{equation}
Therefore from (\ref{geodesics}) we have
\begin{equation}
    \frac{1}{2}\,(\frac{dr}{d\lambda})^2+V_{eff}=\frac{1}{2}\,E^2,
    \label{geodesics2}
\end{equation}
where the effective potential is given by
\begin{equation}
    V_{eff} (L,r)=\frac{f (r)}{2}\,[\frac{L^2}{r^2}+\epsilon].
    \label{potential}
\end{equation}
This potential equals 0 at the surface of the black holes.

According to the classical black hole (BH) thermodynamics \cite{Bek,Bek2,SWH,SWH2}, the surface gravity of BH, $k$, is connected with the temperature $T_h$ known as Hawking temperature, given by,
\begin{equation}
    T_h=\frac{k}{2\,\pi}.
    \label{temp}
\end{equation}
In our case we have the black holes temperature
\begin{equation}
    T_{h} (r_0)=\frac{2\,b^2\,r_0}{2\,\pi}=\frac{\sqrt{\alpha}}{\pi}\,[-\frac{\Lambda}{3}]^{\frac{3}{4}}.
    \label{temp2}
\end{equation}

Thus from above analysis we have seen that the solution (\ref{1}) in this case represent a non-rotating cylindrical symmetry and static charged singular black hole solution with anisotropic fluid as the stress-energy tensor violating the weak energy condition (WEC) in the backgrounds of anti-de Sitter (AdS) spaces.

\vspace{0.1cm}
\begin{center}
{\bf Case 3} : $\beta\neq 0$,\quad $\alpha\neq 0$,\quad $b^2=0$,\quad $\gamma=0$,\quad $\delta\neq 0$.
\end{center}
\vspace{0.1cm}

The metric under this case is given by 
\begin{equation}
    ds^2=-f(r)\,dt^2+\frac{dr^2}{f(r)}+r^2\,(d\phi^2+dz^2),
    \label{case3-metric}
\end{equation}
where the function $f(r)$ is given by
\begin{equation}
f(r)=(\frac{\alpha^2}{r^2}+\frac{2\,\beta}{r}+\delta\,r).
\label{nn1}
\end{equation}
The Krestchmann scalar given by
\begin{equation}
R_{\mu\nu\rho\sigma}\,R^{\mu\nu\rho\sigma}=\frac{8}{r^8}\,[7\,\alpha^4+12\,r\,\alpha^2\,\beta+6\,r^2\,\beta^2-r^3\,\alpha^2\,\delta+r^6\,\delta^2]
\label{nn2}
\end{equation}
\textcolor{blue}{is singular at $r=0$ and tends to zero, as $r\rightarrow \infty$.}

The non-zero components of the Einstein's tensor are
\begin{equation}
G^{0}_{0}=G^{1}_{1}=-\frac{\alpha^2}{r^4}+\frac{2\,\delta}{r},\quad G^{2}_{2}=G^{3}_{3}=\frac{\alpha^2}{r^4}+\frac{2\,\delta}{r}.
\label{nn3}
\end{equation}
Considering the stress-energy tensor the electromagnetic field given by 
\begin{equation}
-T^{0}_{0}=-T^{1}_{1}=T^{2}_{2}=T^{3}_{3}=\frac{B^2}{\kappa\,r^4},
\label{nn4}
\end{equation}
where we have chosen the electromagnetic field tensor $F_{32}=-F_{23}=B^1=B$ and others are zero. The stress-energy tensor of aniostropic fluid is given by (\ref{anisotropic}).

Therefore for zero cosmological constant, from the field equations $G_{\mu\nu}=\kappa\,T_{\mu\nu}=\kappa\,(T_{\mu\nu}(\mbox{e.m.})+T_{\mu\nu}(\mbox{f}))$, we get
\begin{equation}
B=\alpha,\quad -\kappa\,\rho=\kappa\,p_{r}=\frac{2\,\delta}{r},\quad \kappa\,p_{t}=-\frac{\delta}{r}.
\label{nn6}
\end{equation}
\textcolor{blue}{From above it is clear that the the physical parameters $\rho, p_{r},p_{t}$ are singular at $r=0$ and tends to zero at $r \rightarrow \infty$.} The matter-energy tensor the anisotropic fluid satisfy the following energy conditions provided $\delta>0$
\begin{eqnarray}
&&WEC\quad : \quad\quad\quad \rho<0,\nonumber\\
&&WEC_{r}\quad :\quad\quad \rho+p_{r}=0,\nonumber\\
&&WEC_{t}\quad :\quad\quad \rho+p_{t}<0,\nonumber\\
&&SEC\quad\quad:\quad\quad \rho+\sum_{i} p_{i}<0.
\label{energy-1}
\end{eqnarray}
And for $\delta<0$ we have
\begin{eqnarray}
&&WEC\quad : \quad\quad\quad \rho>0,\nonumber\\
&&WEC_{r}\quad :\quad\quad \rho+p_{r}=0,\nonumber\\
&&WEC_{t}\quad :\quad\quad \rho+p_{t}>0,\nonumber\\
&&SEC\quad\quad:\quad\quad \rho+\sum_{i} p_{i}>0.
\label{energy-2}
\end{eqnarray}

A cylindrically symmetric space-time locally is defined by the existence of two commuting, spacelike Killing vectors, whose one orbits are closed and other is open. According to the definition in Ref. \cite{JC}, a space-time $(M, g)$ is cylindrically symmetric if and only if it admits a $G_2$ on $S_2$ group of isometries containing an axial symmetry (see also, Refs.\cite{MM,JC2,SH}). For the space-time (\ref{case3-metric}), the metric tensor is independent of the coordinates $(\phi, z)$ so that the Killing vectors are $(\xi_{(\phi)}, \xi_{(z)})=(\partial_{\phi}, \partial_{z})$. The existence of axial symmetry about $z$-axis (since $|g_{\phi\phi}|\rightarrow 0$, as $r\rightarrow 0$) implies that the orbits of one of these Killing vectors, namely, $\partial_{\phi}$ are closed but those of the other ($\partial_{z}$) is open. In addition, these spacelike Killing vectors generates a two-dimensional isometry group $G_2$ and they commute. Here the assumption on the existence of two-surfaces ($S_2$) orthogonal to the group orbits is not necessary for the definition of cylindrical symmetry. So in the astrophysical context, the orthogonal transitivity in the definition of cylindrical symmetry is removed and we assume the space-time is cylindrically symmetric.

The norm of the spacelike Killing vectors $(\partial_{\phi}, \partial_{z})$ are invariant, namely, the circumferential radius
\begin{equation}
\zeta=\sqrt{|\xi_{(\phi)\mu}\,\xi_{(\phi)\nu}\,g^{\mu\nu}|},
\label{ab1}
\end{equation}
and the specific length
\begin{equation}
l=\sqrt{|\xi_{(z)\mu}\,\xi_{(z)\nu}\,g^{\mu\nu}|}.
\label{ab2}
\end{equation}
The gravitational energy per unit specific length in cylindrical symmetry system (also known as C-energy) is defined as \cite{Thor,Th}
\begin{equation}
{\bf U}=\frac{1}{8}\,\left(1-\frac{1}{4\,\pi^{2}}\,l^{-2}\,\nabla^{\mu} {\bf A}\,\nabla_{\mu} {\bf A}\right),
\label{ab3}
\end{equation}
where ${\bf A}=\zeta\,l$ is the areal radius (area of two-dimensional cylindrical surface where the cylindrical space-time lies), and ${\bf U}$ is the C-energy scalar.

For the metric (\ref{case3-metric}), $\zeta=r$ and $l=r$ so that area of the two-dimensional cylindrical surface ${\bf A}=r^2$. Hence the C-energy scalar ${\bf U}$ is
\begin{equation}
{\bf U}=\frac{1}{8}\,[1-\frac{f(r)}{\pi^{2}}].
\label{ab4}
\end{equation}
The C-energy is defined in terms of the C-energy flux vector $P^{\mu}$ which satisfies the conservation law $\nabla_{\mu}\,P^{\mu}=0$. The C-energy flux vector $P^{\mu}$ is defined by
\begin{equation}
P^{\mu}=2\,\pi\,\epsilon^{\mu\nu\rho\sigma}\,{\bf U}_{,\nu}\,\xi_{(z)\rho}\,\xi_{(\phi)\sigma},
\label{flux}
\end{equation}
where $\epsilon^{\mu\nu\rho\sigma}$ is the Levi-Civita skew tensor, and the C-energy flux vector is
\begin{equation}
P^{\mu}=-\frac{r^4\,f' (r)}{4\,\pi}\,(1,0,0,0).
\label{ab5}
\end{equation}
The C-energy density measured by an observer whose world-line has a tangent vector $u^{\mu}$ is
\begin{equation}
{\bf E}(r)=-P^{\mu}\,u_{\mu}=\frac{r^4\,f' (r)\,\sqrt{f (r)}}{4\,\pi},\quad \mbox{where}\quad u^{\mu}=\frac{1}{\sqrt{f (r)}}\,\delta^{\mu}_{t}.
\label{flux2}
\end{equation}

\underline{\bf Sub-case} (i): If $\beta=0=\delta$.

The metric in this case reduces to 
\begin{equation}
ds^2=-\frac{\alpha^2}{r^2}\,dt^2+\frac{r^2}{\alpha^2}\,dr^2+r^2\,(d\phi^2+dz^2).
\label{28}
\end{equation}
By applying the transformations $t\rightarrow v-\frac{r^3}{3\,\alpha^2}$ into the above metric, one will get
\begin{equation}
ds^2=-\frac{\alpha^2}{r^2}\,dv^2+2\,dv\,dr+r^2\,(d\phi^2+dz^2).
\label{29}
\end{equation}
The above solution represents a Petrov type D static and cylindrical symmetry model with non-null electromagnetic field possesses a naked singularity.

The Lagrangian of metric (\ref{28}) is
\begin{equation}
{\it L}=\frac{1}{2}\,[-\frac{\alpha^2}{r^2}\,\dot{t}^2+\frac{r^2}{\alpha^2}\,\dot{r}^2+r^2\,(\dot{\phi}^2+\dot{z}^2)].
\label{33}
\end{equation}
There are three constant of motions corresponding to three cyclic coordinates $t$, $\phi$ and $z$. These are
\begin{equation}
\dot{t}=\frac{r^2}{\alpha^2}\,E,\quad p_{\phi}=r^2\,\dot{\phi},\quad p_{z}=r^2\,\dot{z}.
\label{34}
\end{equation}
Substituting (\ref{34}) into (\ref{33}) for null geodesics, we have
\begin{equation}
    \frac{\dot{r}}{\sqrt{1-\frac{a^2}{r^4}}}=E,\quad a^2=\alpha^2\,(p^{2}_{\phi}+p^{2}_{z}).
    \label{case3-1}
\end{equation}
For the radial null geodesic $\dot{\phi}=0$ in the constant $z-plane$, we have from (\ref{33})
\begin{equation}
\dot{r}=\frac{\alpha^2}{r^2}\,\dot{t}=E\Rightarrow r(\lambda)=E\,\lambda+c_1,
\label{35}
\end{equation}
where $c_1$ is arbitrary constant. Therefore from (\ref{34}) for time coordinate
\begin{equation}
\dot{t}=\frac{E}{\alpha^2}\,(E\,\lambda+c_1)^2\Rightarrow t(\lambda)=\frac{1}{3\,\alpha^2}\,(E\,\lambda+c_1)^3.
\label{36}
\end{equation}
From Eq. (\ref{35}) and (\ref{36}) it is clear that the radial null geodesics path $t$ and $r$ are bounded for finite value of the affine parameter $\lambda$, and thus the presented space-time is radially null geodesically complete but incomplete for non-radial null geodesics.

Consider a congruence of radial null geodesics having the tangent vector $K^{\mu}(r)=(K^{t},K^{r},0,0)$, where $K^{t}=\frac{dt}{d\lambda}$ and $K^{r}=\frac{dr}{d\lambda}$. The geodesic expansion is defined by \cite{SINGH}
\begin{equation}
\boldsymbol{\Theta}=K^{\mu}_{\,\,;\,\mu}=\frac{1}{\sqrt{-g}}\,\frac{\partial}{\partial\,x^{i}}\,(\sqrt{-g}\,K^{i}).
\label{37}
\end{equation}
For the metric (\ref{28}) we get  
\begin{equation}
\boldsymbol{\Theta}=\frac{\partial K^{r}}{\partial r}+\frac{K^{r}}{\sqrt{-g}}\,(\sqrt{-g})^{'},
\label{38}
\end{equation}
where prime denotes derivative w. r. t. r. We proceed by noting that
\begin{eqnarray}
\frac{dK^{r}}{d\lambda}&=&\frac{\partial K^{r}}{\partial r}\,\frac{\partial r}{\partial \lambda},\nonumber\\
\frac{dK^{t}}{d\lambda}&=&\frac{\partial K^{t}}{\partial r}\,\frac{\partial r}{\partial \lambda}.
\label{39}
\end{eqnarray}
Therefore the final expression of the geodesic expansion from (\ref{38}) is
\begin{eqnarray}
\boldsymbol{\Theta}(r)&=&\frac{1}{K^{r}}\,\frac{dK^{r}}{d\lambda}+\frac{2}{r}\,K^{r}\nonumber\\
&=&\frac{1}{K^{r}}\,[-\Gamma^{r}_{tt}\,(K^{t})^2-\Gamma^{r}_{rr}\,(K^{r})^2]+\frac{2}{r}\,K^{r}\nonumber\\
&=&-\frac{(K^{t})^2}{K^{r}}\,\Gamma^{r}_{tt}-\Gamma^{r}_{rr}\,K^{r}+\frac{2}{r}\,K^{r}\nonumber\\
&=&\frac{(K^{t})^2}{K^{r}}\,\frac{\alpha^4}{r^5}-\frac{1}{r}\,K^{r}+\frac{2}{r}\,K^{r}\nonumber\\
&=&\frac{2\,E}{r},\quad K^{t}=\frac{r^2}{\alpha^2}\,E\nonumber\\
&>&0,
\label{40}
\end{eqnarray}
which is positive, a fact that indicates the nature of singularity which is formed due to scalar curvature is naked. One can easily check that the naked singularity (NS) which is formed due to the scalar curvature satisfies neither the {\it strong curvature condition} developed by Tipler \cite{Tipler} (see also, \cite{Clark}) nor the {\it limiting focusing condition} developed by Krolak \cite{Kro}. These two criteria are as follows :
\begin{equation}
\lim_{\lambda\rightarrow 0} \lambda^2\,R_{\mu\nu}\,\frac{dx^{\mu}}{d\lambda}\,\frac{dx^{\nu}}{d\lambda}\neq 0(>0),
\label{41}
\end{equation}
where $\frac{dx^{\mu}}{d\lambda}$ is the tangent vector field to the radial null geodesics. And
\begin{equation}
\lim_{\lambda\rightarrow 0} \lambda\,R_{\mu\nu}\,\frac{dx^{\mu}}{d\lambda}\,\frac{dx^{\nu}}{d\lambda}\neq 0.
\label{42}
\end{equation}
Therefore, the analytical extension of the space-time through the singularity is possible.

The C-energy density in this sub-case is given by
\begin{equation}
    E (r)=-\frac{\alpha^3}{2\,\pi}<0
    \label{ccc1}
\end{equation}
which may positive or negative depending on the sign of $\alpha$.

\vspace{0.1cm}
\underline{\bf Sub-case} (ii): If $\beta=0=\alpha$.
\vspace{0.1cm}

The metric in this case reduces to
\begin{equation}
ds^2=-\delta\,r\,dt^2+\frac{1}{\delta\,r}\,dr^2+r^2\,(d\phi^2+dz^2).
\label{46}
\end{equation}
And doing transformation $t\rightarrow v-\frac{1}{\delta}\,\mbox{ln} r$ into the above metric, one will get
\begin{equation}
ds^2=-\delta\,r\,dv^2+2\,dv\,dr+r^2\,(d\phi^2+dz^2).
\label{47}
\end{equation}
This solution represents a cylindrical symmetry and conformally flat static solution of only anisotropic fluid (\ref{nn6}). It is worth mentioning the conformally flat condition for static and cylindrically symmetric case with anisotropic fluids was obtained by L. Herrera {\it et al} \cite{Herr1}. They had shown that any conformally flat and cylindrically symmetric static source cannot be matched through Darmois conditions to the Levi-Civita space-time, satisfying the regularity conditions. Furthermore, all static, cylindrical symmetric solutions (conformally flat or not) for anisotropic fluids have been found in \cite{Herr2}. Very recently, a conformally flat cylindrical symmetry and static anisotropic solution of the field equations with naked singularities appeared in \cite{Faiz6}. Therefore the study static and cylindrical symmetry space-time (\ref{46}) is a special case of these known solutions.

The C-energy density in this sub-case is given by
\begin{equation}
    E (r)=\frac{\delta^{\frac{3}{2}}\,r^{\frac{9}{2}}}{4\,\pi}.
    \label{ccc2}
\end{equation}

\vspace{0.1cm}
\underline{\bf Sub-case} (iii) : For $\delta=0=\alpha$,
\vspace{0.1cm}

The metric in this case reduces to
\begin{equation}
ds^2=-\frac{2\,\beta}{r}\,dt^2+\frac{r}{2\,\beta}\,dr^2+r^2\,(d\phi^2+dz^2).
\label{48}
\end{equation}
And by doing transformation $t\rightarrow v-\frac{r^2}{4\,\beta}$ into the metric, one will get
\begin{equation}
ds^2=-\frac{2\,\beta}{r}\,dv^2+2\,dv\,dr+r^2\,(d\phi^2+dz^2).
\label{4444}
\end{equation}
The metric (\ref{50}) represents a cylindrical symmetry type D vacuum solution of the field equations since $R_{\mu\nu}=0$. This type D vacuum solution possess a naked curvature singularity at $r=0$ not covered by an event horizon.

We determine the fourteen scalar invariants of the Riemann-Christoffel curvature tensor for the vacuum spacetime from the list provided in \cite{Harv} which is based on the one in \cite{Narli}. Of these fourteen, only four do not vanish identically for a vacuum spacetime (Ricci flat). These are (using the notation of \cite{Harv,Narli})
\begin{eqnarray}
&&J_1=A_{hijk}\,g^{hj}\,g^{ik}\quad,\nonumber\\
&&J_2=B_{hijk}\,g^{hj}\,g^{ik}\nonumber\quad,\\
&&J_3=A_{ab}^{\quad cd}A_{cd}^{\quad ab}-\frac{1}{2}\,J^2_1\quad,\\
\label{case3-11}
&&J_4=A_{ab}^{\quad cd}B_{cd}^{\quad ab}-\frac{5}{12}\,J_1\,J_2\quad,\nonumber
\end{eqnarray}
where the tensors $A_{hijk}$ and $B_{hijk}$ are obtained from the Weyl tensor $C$ (which is identical to the Riemann tensor in empty space) via
\begin{eqnarray}
&&A_{hijk}=C_{hipq}\,C_{rsjk}\,g^{pr}\,g^{qs}\nonumber\\
&&B_{hijk}=C_{hipq}\,A_{rsjk}\,g^{pr}\,g^{qr}.
\label{case3-111}
\end{eqnarray}
For the metric (\ref{48}), we find that
\begin{equation}
J_1=J_2=\frac{48\,\beta^2}{r^6},\quad J_3=-\frac{576\,\beta^4}{r^{12}}\quad \mbox{and} \quad J_4=-\frac{384\,\beta^2}{r^{12}}.
\label{case3-12}
\end{equation}
From above analysis with the Kretschmann scalar, we have seen that the curvature invariants diverge at $r=0$ not covered by an event horizon. 

The C-energy density in this sub-case is given by
\begin{equation}
    E (r)=-\frac{(2\,\beta\,r)^{\frac{3}{2}}}{4\,\pi}<0
    \label{ccc3}
\end{equation}
\textcolor{blue}{By doing similar analysis as done in sub-case (i) earlier}, one can show that the singularity which is formed due to scalar curvature is naked which satisfies neither the {\it strong curvature condition} nor the {\it limiting focusing condition}. Recently, the author with other \cite{Faiz5} constructed a cylindrical symmetric type D vacuum solution of the field equations which is free-from causality possesses a naked singularity.

\vspace{0.1cm}
\begin{center}
{\bf Case 4} : $\beta=0$,\quad $\alpha=0$,\quad $b^2 \neq 0$,\quad $\delta=0$,\quad $\gamma\neq 0$.
\end{center}
\vspace{0.1cm}

The metric is given by
\begin{equation}
ds^2=-b^2\,r^2\,dt^2+\frac{1}{b^2}\,\frac{dr^2}{r^2}+r^2\,(e^{-2\,\gamma\,t}\,d\phi^2+e^{2\,\gamma\,t}\,dz^2).
\label{mmnn}
\end{equation}
The Krestchamnn scalar given by
\begin{equation}
R^{\mu\nu\rho\sigma}\,R_{\mu\nu\rho\sigma}=24\,b^4+\frac{12\,\gamma^4}{b^4\,r^4}-\frac{8\,\gamma^2}{r^2}
\label{49}
\end{equation}
\textcolor{blue}{is singular at $r=0$} not clothed and thus a naked singularity is formed. At spatial infinity along the radial direction, that is, at $r\rightarrow \infty$, the curvature scalar $R^{\mu\nu\rho\sigma}\,R_{\mu\nu\rho\sigma}\rightarrow 24\,b^4$ which indicates that the solution is asymptotically de-Sitter or anit-de Sitter spaces. One can easily check the solution is of Petrov type D metric with naked singularities.

Consider the stress-energy tensor the perfect fluid (\ref{10}), from the field equations (\ref{1}) and (\ref{5}), we get
\begin{equation}
\Lambda=-3\,b^2,\quad \kappa\,\rho=\kappa\,p=-\frac{\gamma^2}{b^2\,r^2}, \quad \Lambda<0 \quad \mbox{or} >0.
\label{50}
\end{equation}

There are two possible sub-cases arise :

\vspace{0.1cm}

\underline{Sub-case} (i): For $b^2<0$, that is, $\Lambda>0$, $t$ is space-like and $r$ is a time-like coordinate. The matter-energy sources the stiff perfect fluid satisfies the energy condition. The physical parameter $\rho (=p)$ is singular at $r=0$ and tends to zero at spatial infinity along the radial distances {\it i. e.,} $r\rightarrow \infty$. Thus the non-static type D solution corresponds to a stiff perfect fluid in the backgrounds of de-Sitter (dS) spaces with naked singularities.

\vspace{0.1cm}

\underline{Sub-case} (ii): For $b^2>0$, that is, $\Lambda<0$, $t$ is time-like and $r$ is a space-like coordinate. In that case, the stiff perfect fluid violates the weak energy condition. Therefore the non-static type D solution corresponds to a stiff fluid in the background of anti-de Sitter (AdS) spaces with naked singularities violating the weak energy condition.

Noted that if one takes $\gamma=0$, then the static space-time represent a de-Sitter or anti-de Sitter spaces, depending on the signs of the cosmological constant $\Lambda$ whose global structure were studied extensively in \cite{Hawking}.

\vspace{0.1cm}
\begin{center}
{\bf Case 5} : $\beta=0$,\quad $\alpha \neq 0$,\quad $b^2 \neq 0$,\quad $\gamma=0$,\quad $\delta  \neq 0$.
\end{center}
\vspace{0.1cm}

The function $f(r)$ and the surface gravity $k(r)$ is given by
\begin{equation}
f(r)=(\frac{\alpha^2}{r^2}+b^2\,r^2+\delta\,r),\quad k(r)=(-\frac{\alpha^2}{r^3}+b^2\,r+\frac{\delta}{2}).
\label{case6}
\end{equation}
In that case, the space-time represents a cylindrical symmetric and static solution with naked singularities which we discuss graphically {\it i. e.} one can see that no horizon exists for specific values of the parameters. Figure 1 indicates that naked singularities exist for positive values of $\alpha$ and $\delta$ with both de-Sitter and anti de-Sitter spaces. Figure 2 indicates that naked singularities exist for negative values of $\alpha$ and positive values of $\delta$ with both de-Sitter and anti de-Sitter spaces.
 
\begin{figure*}[tbp]
\centering
\includegraphics[width=6.0cm,height=5.0cm]{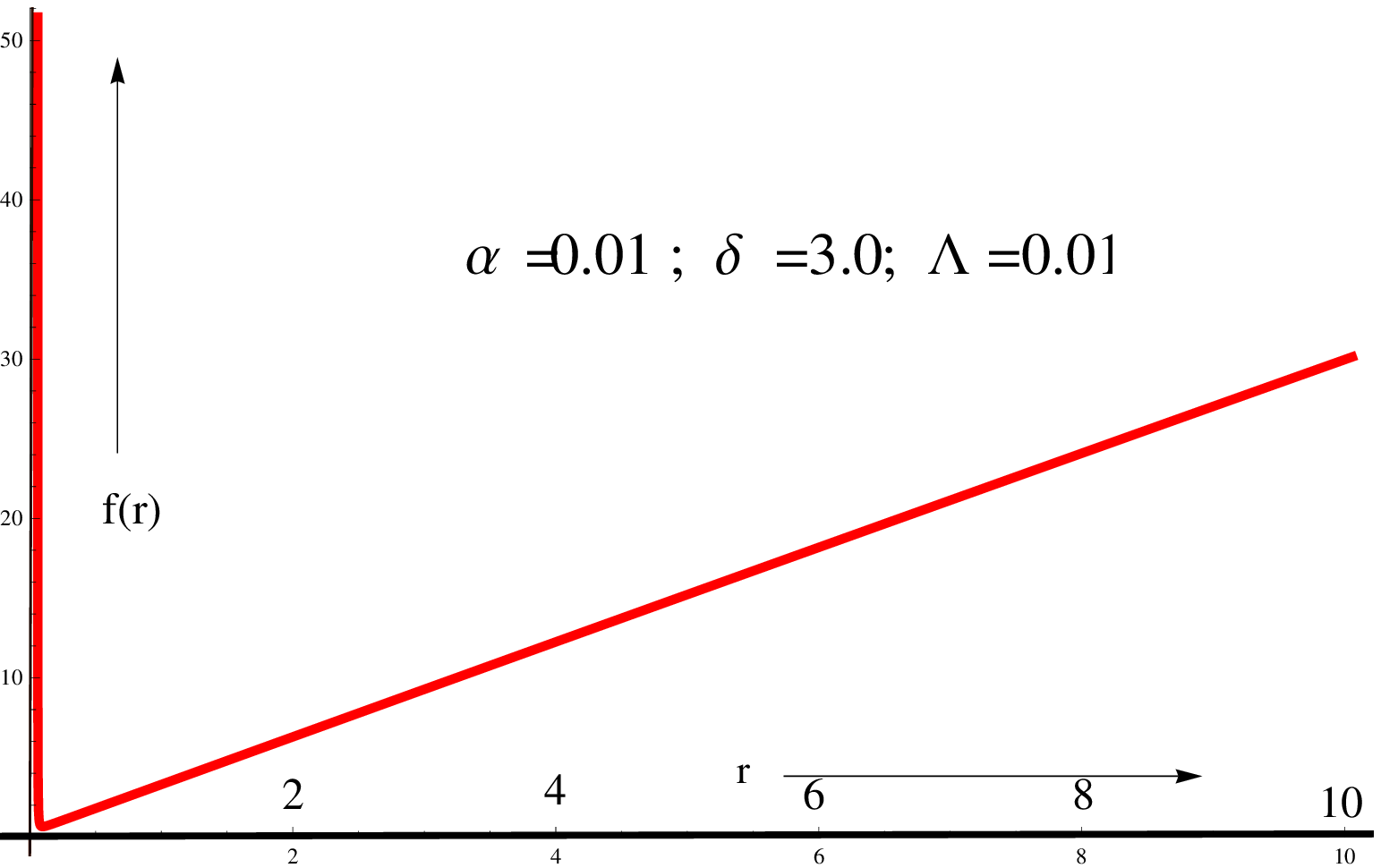}
\hspace{0.5cm}
\includegraphics[width=6.0cm,height=5.0cm]{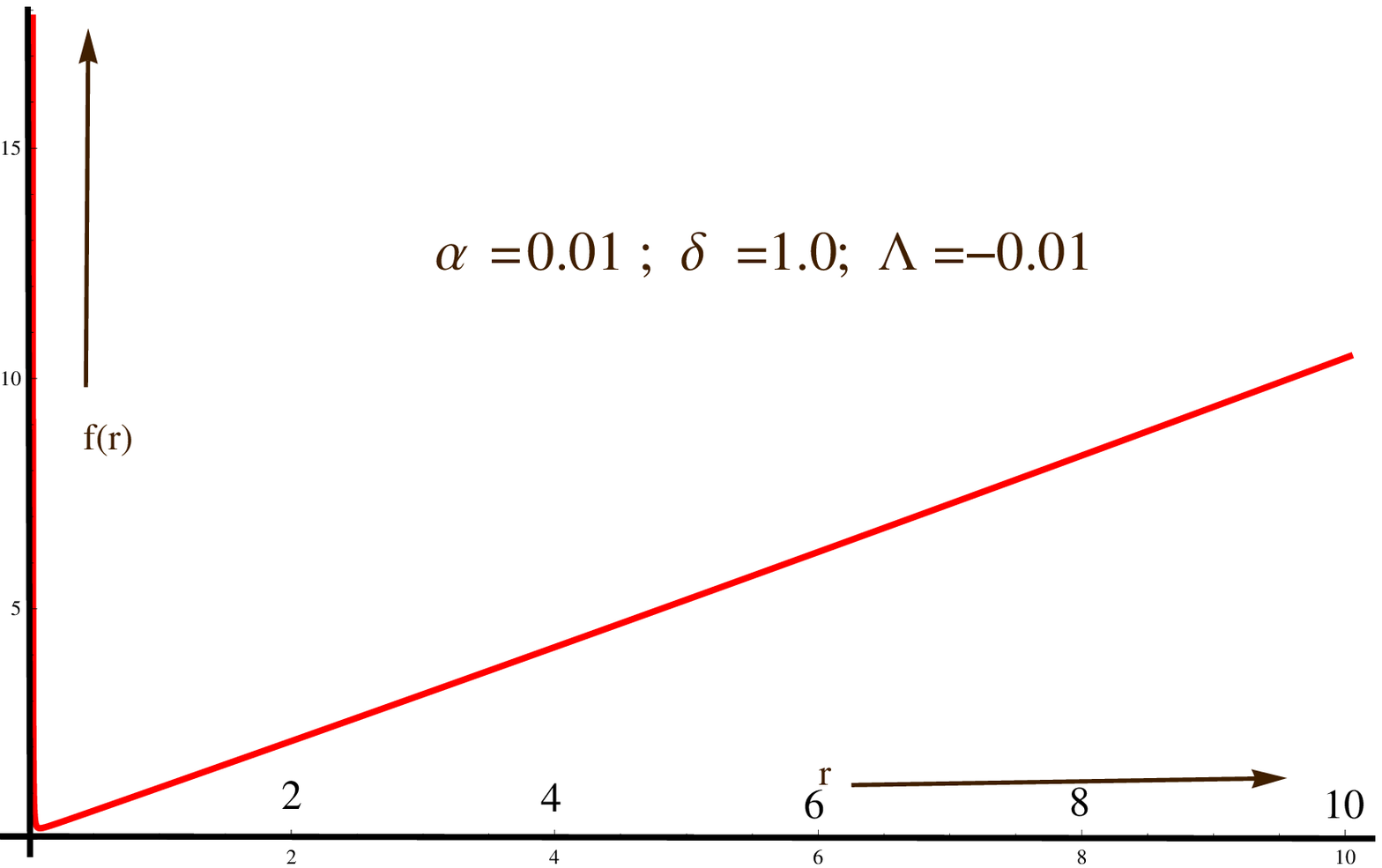}
\caption{Variation of f(r) with respect to r for positive values of $\alpha, \delta$. (left panel) de-Sitter spaces (right panel ) Anti de-Sitter spaces.}
\end{figure*}
\begin{figure*}[tbp]
\centering
\includegraphics[width=6cm,height=5cm]{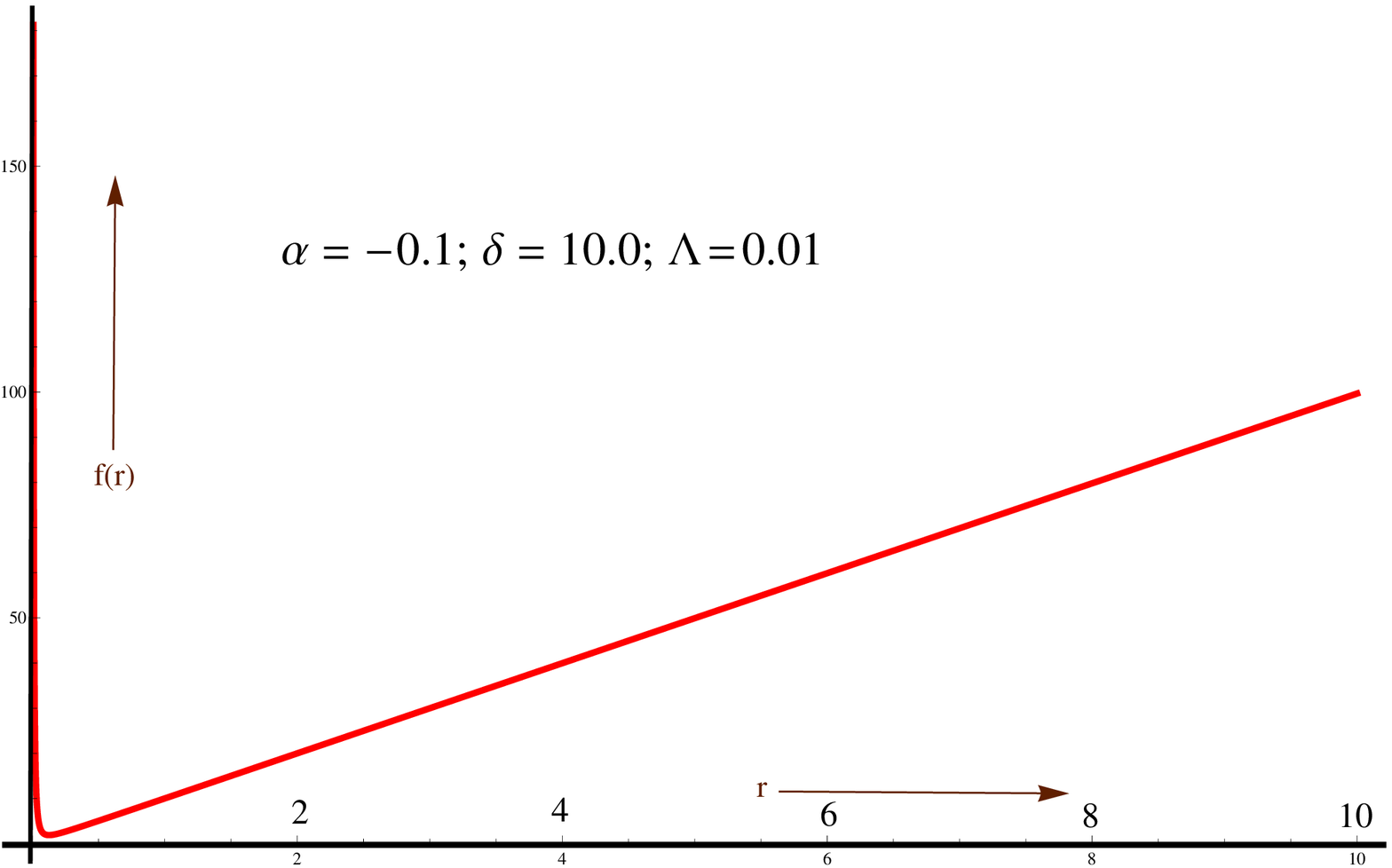}
\hspace{0.5cm}
\includegraphics[width=6cm,height=5cm]{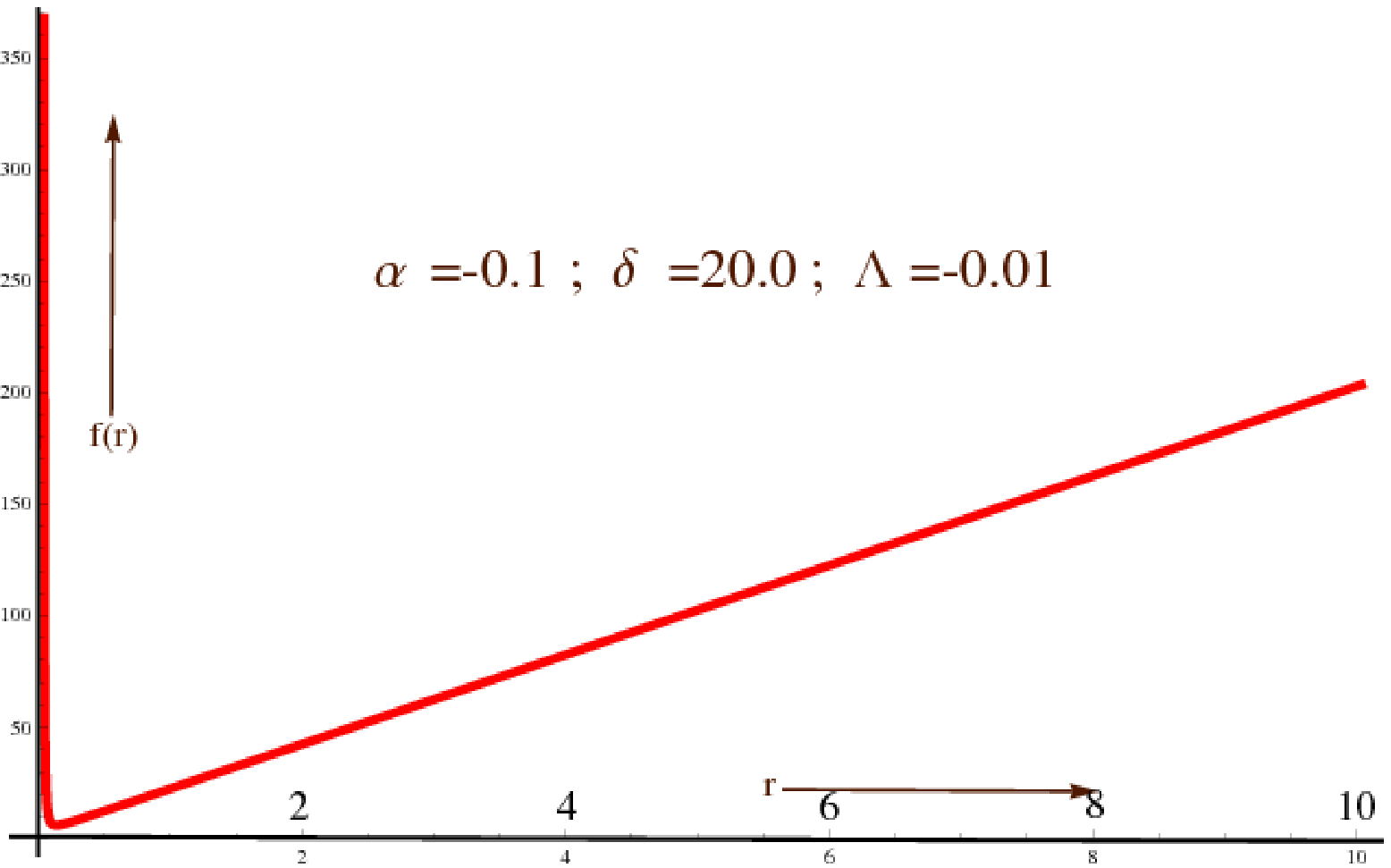}
\caption{Variation of f(r) with respect to r for negative values of $\alpha$ and positive values of $\delta$. (left panel) de-Sitter spaces (right panel )  Anti de-Sitter spaces. }
\end{figure*}

One can easily check that the stress-energy tensor corresponds to non-null electromagnetic field coupled with anisotropic fluid. The different physical parameters are as follows:
\begin{eqnarray}
B&=&\alpha>0,\quad \Lambda=-3\,b^2 < 0\quad \mbox{or}\quad >0,\nonumber\\
\kappa\,\rho&=&-\frac{2\,\delta}{r},\quad \kappa\,p_{r}=\frac{2\,\delta}{r},\quad \kappa\,p_{t}=\frac{\delta}{r}.
\label{51}
\end{eqnarray}
The ansiotropy difference along the radial and tangential directions, respectively are
\begin{equation}
\kappa\,(\rho-p_{r})=-\frac{4\,\delta}{r},\quad \kappa\,(\rho-p_{t})=-\frac{3\,\delta}{r}.
\label{52}
\end{equation}
Therefore the static solution corresponds to an anisotropic fluid coupled with non-null electromagnetic field in the backgrounds of de-Sitter and anti-de Sitter spaces with naked singulrities.

\section{Conclusions}

The General Theory Relativity (GTR) is the successful theory not only because of the predictions of black holes but also its capability of predicting its shortcomings. One of the apparent shortcomings of General Relativity is that it predicts the existence of singularities, space-time geometrical points where curvature becomes infinite and laws of physics are no longer working. Interestingly the Penrose-Hawking singularity theorems do not say much about geometrical locations of singularities \cite{Hawking}. Since the introduction of cosmic censorship conjecture by Penrose, many endeavours attempted to argue in favour (please see, \cite{RMW}) or against the weak and strong versions of cosmic censorship conjecture. None came with conclusive definitive proof whether naked singularities could or could not physically exist. On the contrary, there are a number of spherical and non-spherical gravitational collapse solution with naked singularities exist (For a review, see, \cite{Jos1}). For regular or non-singular black hole solution, Bardeen proposed the first static and spherically symmetric regular black hole solution \cite{Bardeen}. After that several authors investigated this type black hole solution. For a comprehensive review of this type black hole solution, see \cite{Ansoldi}. In the present work, we are mainly interested on singular black hole solution where the central singularity is covered by an event horizon.

A four dimensional non-static or static space-time with the matter-energy sources the stiff fluid,  anisotropic fluid and an electromagnetic field, is analyzed here. The space-time represents naked singularity models and/or singular black hole solutions depending on the various parameters conditions on the metric functions in the background of de-Sitter (dS) or anti de-Sitter (AdS) spaces. In {\it Case 1}, we have presented a generalization of cylindrical symmetric and static Lemos black hole solution into non-static one with a perfect fluid satisfying the weak energy condition (WEC) within the black hole region. In {\it Case 2}, a cylindrical symmetric and static black hole solution with a negative cosmological constant and anisotropic fluid as the stress-energy tensor, is discussed in details. In {\it Case 3}, type D cylindrical symmetric and non-static solutions of non-null electromagnetic field and anisotropic fluid as the stress-energy tensor with a naked curvature singularity, is presented. There we have discussed three sub-cases by (i)---(iii) of the space-time. In sub-case (i), we have constructed a type D cylindrical symmetric and static space-time of non-null electromagnetic field only which possesses a naked curvature singularity. We have shown that the space-time is radially null geodesically complete. Also we have shown there that the singularity which is formed due to scalar curvature (or the Krestchmann scalar) is naked and satisfies neither the strong curvature condition nor the limiting focusing condition. In sub-case (ii), a cylindrical symmetric and conformally flat type $O$ solution of anisotropic fluid only with a naked curvature singularity, is presented. It is worth mentioning the conformally flat condition for static and cylindrically symmetric case with anisotropic fluids was obtained by L. Herrera {\it et al} \cite{Herr1}. Furthermore, all static, cylindrical symmetric solutions (conformally flat or not) for anisotropic fluids have been found in \cite{Herr2}. Thus cylindrical symmetric and conformally flat static space-time discussed in sub-case (ii) is a special case of these known solutions. In sub-case (iii), we constructed a cylindrical symmetric and static type D vacuum space-time with a naked curvature singularity. In {\it Case 4}, type D cylindrical symmetric and non-static solution of stiff perfect fluid in the backgrounds of de-Sitter (dS) and anti-de Sitter (AdS) spaces with a naked curvature singularity, is presented. In {\it case 5}, cylindrical symmetric and static solution of anisotropic fluid coupled with a non-null electromagnetic field in the backgrounds of de-Sitter (dS) and anti-de Sitter (AdS) spaces with a naked singulrity, is presented. We have plotted graphs (figs. 1 and 2) showing variation of the function $f(r)$ against $r$ in the backgrounds of de-Sitter and anti-de Sitter spaces by choosing suitable parameters $\alpha$ and $\delta$.

\section*{Acknowledgement} I would like to thank the anonymous kind referee(s) for his/her valuable comments and suggestions. I am also very thankful to Prof. Luis Herrera, Universidad de Salamanca, Spain, for fruitful discussion and valuable suggestions.

\end{document}